\documentclass[useAMS,usenatbib]{mn2e}
\usepackage{subfig}
\usepackage{graphicx}
\usepackage{rotating}
 \usepackage{float}
\usepackage{amsmath}

\newcommand{\Mpc}{$h^{-1}$ Mpc}
\newcommand{\Msun}{$h^{-1}$ M$_{\odot}$}
\newcommand{\dn}{$\delta_{NN}$}
\newcommand{\df}{$\delta_{FA}$}

\newcommand{\dps}{$\delta^{proj}_{spec}$}
\newcommand{\dpp}{$\delta^{proj}_{photo}$}
\newcommand{\dtD}{$\delta^{3D}_{real}$}
\newcommand{\kms}{km $s^{-1}$}

\begin{document}

\title[Measures of Galaxy Environment III]{Measures of Galaxy Environment - III. Difficulties in identifying proto-clusters at $z \sim 2$}
\author[G. M. Shattow et al.]{Genevieve M. Shattow,$^1$ Darren J. Croton,$^1$ Ramin A. Skibba,$^2$ \newauthor
Stuart I. Muldrew,$^3$ Frazer R. Pearce,$^3$ and Ummi Abbas$^4$\\
$^1$Swinburne University of Technology, Hawthorn, VIC 3122, Australia\\
$^2$Center for Astrophysics and Space Sciences, Department of Physics, University of California, 9500 Gilman Dr., San Diego, CA 92093, USA \\
$^3$School of Physics \& Astronomy, University of Nottingham, University Park, Nottingham NG7 2RD, UK\\
$^4$INAF, Osservatorio Astronomico di Torino, Strada Osservatorio 20, 10025 Pino Torinese, Italy}

\maketitle

\begin{abstract}

Galaxy environment is frequently discussed, but inconsistently defined. It is especially difficult to measure at high redshift where only photometric redshifts are available. With a focus on early forming proto-clusters, we use a semi-analytical model of galaxy formation to show how the environment measurement around high redshift galaxies is sensitive to both scale and metric, as well as to cluster viewing angle, evolutionary state, and the availability of either spectroscopic or photometric data. We use two types of environment metrics (nearest neighbour and fixed aperture) at a range of scales on simulated high-z clusters to see how ``observed" overdensities compare to ``real" overdensities. We also ``observationally" identify $z = 2$ proto-cluster candidates in our model and track the growth histories of their parent halos through time, considering in particular their final state at $z = 0$. Although the measured environment of early forming clusters is critically dependent on all of the above effects (and in particular the viewing angle), we show that such clusters are very likely ($\ga 90\%$) to remain overdense at $z = 0$, although many will no longer be among the most massive. Object to object comparisons using different methodologies and different data, however, require much more caution.

\end{abstract}


\section{Introduction}\label{Introduction}

Quantifying galaxy environment and its influence on galaxy evolution is crucial to understanding both the individual and statistical properties of a galaxy population. Luckily, environment is an easily measurable quantity in modern extragalactic surveys and hence has become a popular target for study. Despite this, environment is not uniquely defined. Typically, different authors will employ variations on common (but distinct) environment methodologies, optimised for the particular data on hand as well as the galaxy property of interest. In addition, different ways to measure galaxy environment are differently sensitive to the data, sometimes in unexpected ways. This is especially true at high redshift where the data quality becomes difficult and costly to maintain. This can make comparisons between results in the literature problematic, something we wish to address in the current work.

Broadly speaking, environment has been shown to correlate strongly with galaxy properties such as morphology \citep{Dressler1980, Postman1984}, colour \citep{Hogg2004, Wilman2010}, luminosity \citep{Norberg2001, Norberg2002, Blanton2005, Croton2005}, structure and shape \citep{Blanton2005, Skibba2012}, and clustering \citep{Abbas2006}, among others. In particular, the processes that affect a galaxy in denser environments are often lacking in less dense environments, e.g. ram pressure stripping, or the removal of gas from a galaxy as it travels through the intracluster medium \citep{Gunn1972}; harassment, or the rapid interactions with other galaxies in a dense environment \citep{Farouki1981, Moore1996}; assembly bias, or the effect of the formation history on the galaxy \citep{Gao2005, Croton2007}, etc.

Arguably, the two most popular methods to quantify individual galaxy environment are the distance to the Nth nearest neighbour \citep{Dressler1980, Baldry2006, Brough2011} and the number of neighbouring galaxies found within a fixed aperture \citep{Hogg2003, Kauffmann2004, Croton2005}. Other probes of environment and environment classification include the small and large-scale clustering of galaxy populations \citep{Peebles1973, Zehavi2002, Zehavi2005}, the identification of voids and under-dense regions in which lone groups sit \citep{Hoyle2005}, shape statistics that quantify the topology of the cosmic web \citep[e.g. sheet, filament, and cluster;][]{Dave1997}, and estimators of dark matter halo mass (e.g. through velocity dispersion), in which galaxies and galaxy groups sit  \citep{Berlind2006, Yang2007, deLucia2012}. Marked statistics are also a useful probe of scale-dependent feature-environment relations \citep{Sheth2004, Skibba2013}. Scale, in fact, is an important consideration when interpreting the role of environment in galaxy evolution. For example, correlations that might be found by statistics of a 3rd nearest neighbour could be missed by the statistics of a 10th nearest neighbour, and vice-versa.

Although environment measurements can vary significantly, many observed correlations are jointly confirmed using different metrics. For example, \cite{Kauffmann2004} use a 2 \Mpc\ cylindrical aperture with a 500 \kms\ velocity cut and \cite{Poggianti2008} use a 10th nearest projected neighbour measurement with a photometric velocity cut of $\pm 0.1$\ in redshift to find similar relations between density and specific star formation. Clearly this relation holds between these two metrics, but does it continue to hold at larger apertures or with a different value of N? Such results are also often dependent on the population of galaxies used to define the background density. In the first paper in this series, \cite{Muldrew2012} make a detailed comparison of various methods and selections using simulated data and explore the ways in which galaxy properties correlate with the different environment definitions, discussing several of these issues at length.

The importance of environment is often posed as a battle between nature and nurture. However this may be the wrong way to view its role.  \cite{deLucia2012} proposes that the nature vs. nurture argument in galaxy evolution is ill-posed, since the evolution of a galaxy relies on the history of its environment, which might have been drastically different in the past, and not just its current surroundings. The majority of studies of environment cited above have been done in the low-redshift universe, mainly because that is where the best data are available for study. To further explore environmental histories, we need more data at higher redshifts, and accurate galaxy evolution simulations to explore beyond the observations.

Initial exploration at high redshift has focused on overdense regions, as these stand out against the background despite being rare and hence hard to find. Clusters are typically favoured because they are large and bright, and the focus of intersecting galaxy filaments. For example, the first galaxy cluster was recorded by Messier in 1784 \citep{Biviano2000}, whereas the first void was not confirmed until 200 years later, with the Bootes void in 1981 \citep{Kirshner1981}.

There has been a race in the past several years to find the biggest, farthest cluster or proto-cluster, and as a result, there have been several recent discoveries. Large, distant clusters include a $1-4\times 10^{14}M_{\odot}$ (dynamical mass) cluster at z = 1.62 \citep{Papovich2010, Tanaka2010} and a $5.3-8 \times 10^{13}$M$_{\odot}$ cluster at z = 2.07 \citep{Gobat2011}. Even more recently, \cite{Spitler2012} have announced finding three $1-3\times 10^{13}M_{\odot}$ (virial mass) clusters at z = 2.2. The farthest thus far has been a proto-cluster at z$\sim$6 with a mass estimate of $M = 2.9\times 10^{14} M_{\odot}$, but with quite a bit of uncertainty in both the virial mass and overdensity calculations \citep{Toshikawa2012}. There have been many more clusters observed which have not been spectroscopically confirmed yet \citep{Andreon2009, Bielby2010, Trenti2012}.

At increasingly high redshifts observers are limited to measuring overdensities projected on the plane of the sky with either spectroscopic or photometric redshift cuts in the line-of-sight direction. \cite{Haas2012} discuss how spectroscopic redshift space distortions affect the correlation of various environment metrics with halo size in a model at $z = 0$. \cite{Cooper2005} compare the 3D real-space and 2D projected redshift space measurements of galaxies in a mock DEEP2 sample, showing examples of both spectroscopic and photometric redshift cuts. They find that photometric redshift cuts make environment measurements less meaningful. Although there are ways of reconstructing galaxies' photometric redshift distributions using spectroscopic measurements of nearby galaxies \citep[e.g.][]{Kovac2010}, this is not yet common practice. 

Observers are also limited to a single line of sight, which can have a huge impact on both mass \cite[as discussed in][]{Noh2012} and environment measurements (as discussed in this paper). This leads to the question of how reliable and accurate these measurements are. In this paper we address this in three parts, mostly focusing on redshift 2, higher than the studies cited above. First, we explore the difference between projected and actual (i.e. real-space 3D) environments measured with different metrics and on different scales. Second, we examine how projection effects change the estimated environment of massive clusters at z=2. And third, we look at the stability of each environment measure for a particular object with time, asking how often are the most dense proto-clusters found at high redshift still the most dense by z=0.

This paper is organised in the following way: in Section \ref{method} we discuss the simulated data and the methods of measuring environment. In Section \ref{results} we compare environment measures applied to our model galaxy population for individual clusters and cluster populations and follow their evolution. We conclude in Section \ref{discussion}, where we discuss our findings and their implications for observations of high redshift proto-clusters. Throughout we assume a $\Lambda$CDM cosmology, following the parameters of the Millennium Simulation (e.g. $\Omega_m = 0.25, \sigma_8 = 0.9$) \citep{Springel2005}, and a Hubble constant of $H_0 = 100 h$\ km s$^{-1}$ Mpc$^{-1}$.

\section{Methodology}\label{method}

\subsection{Data\label{data}}
\subsubsection{Millennium Simulation \& Galaxy Formation Model\label{millennium}}
We use synthetic data to study galaxy environment at different cosmic epochs. This is necessary because real data is usually incomplete and sometimes unreliable, especially across large redshift ranges. Simulated data, on the other hand, has the attractive property of being both complete and precise, although perhaps not correct, and one can easily link galaxy histories and hence the histories of a galaxy's environment. We take our best available galaxy formation model and dark matter simulation to use as a benchmark to test different environment metrics and their evolution.These are an updated version of the semi-analytic model of \citet{Croton2006} run on the Millennium Simulation \citep{Springel2005}.

The Millennium Simulation follows $N = 2160^3$ particles in a $500^3$ (\Mpc)$^3$ box from redshift  $z=127$ to the present day at  $z=0$. From this 64 snapshots of the simulation's evolution are recorded. A halo finder is then applied in post-processing to link halos of common ancestry across time. This produces the set of halo merger trees for the simulated volume. In all, approximately 25 million halos are identified at $z=0$ and their histories followed back until first identification in the simulation. 

The Millennium Simulation contains only dark matter, and hence we must apply additional post-processing to add a galaxy population. The semi-analytic methodology, first proposed by \cite{White1991}, analytically couples baryonic and dark halo evolution via a system of differential equations. It assumes that every halo hosts a galaxy characterised by a number of baryonic reservoirs, and whose properties evolve based on physical processes thought to be important for galaxy formation: gas accretion and cooling, star formation and stellar population synthesis, galaxy mergers and morphological transitions, and feedback from supernovae and active galactic nuclei \citep{Croton2006, Kitzbichler2007, deLucia2007}. To study galaxy environments we use a updated version of the model described in \cite{Croton2006}, the details of which are not important for this study. We refer the interested reader to Croton et al. (in prep.) for further information.

\subsubsection{The Density Defining Population\label{ddp}}
For very high resolution simulations the number of galaxies available for study can be an embarrassment of riches. In large synthetic universes of tens of millions of objects, such as the one in use here, sensible cuts to the data must be made. This both allows us to mimic more closely the configuration of a real galaxy survey and also to make the processing time of our (sometimes complex) analysis more tractable. 

Following \cite{Croton2005} we construct a sample of background galaxies to be used to define the density contours across the simulation volume. This sample is called the ``density defining population''. When the environment of an object of interest is measured it will always be against the density defining population. To find these galaxies, we map the stellar mass function of the model onto the observed stellar mass functions at each redshift \citep[as found in][]{Perez-Gonzalez2008} and take the galaxies with $M_{stars} > M^* - 1.0$, where $M^*$ ~is the knee of the stellar mass function (a Schechter function). $M^*$ increases from 11.16 to 11.46 (in units of $log_{10}(M_{\odot})$) from redshift $z = 0$ ~to $z = 2$, respectively. This decreases the number of background galaxies from $1.91 \times 10^{6}$\ at $z = 0$\ to $1.85 \times 10^{5}$\ at $z = 2$. Other observationally measured stellar mass functions \citep[e.g.][among others]{Marchesini2009, Ilbert2013}, have very disparate values for the parameters of the Schechter function. Because we use density contrast (see section 2.2) rather than density, the actual parameters in the stellar mass function are less important than the relative background densities at the different redshifts. We choose the lower limit of M*-1 to roughly mimic the range of magnitudes available to observers at higher redshifts. Other stellar mass function fits would therefore require slightly different cuts to achieve the same magnitude range. This is an arbitrary but reasonable and well defined cut. Specific results change with different choices of density defining populations, but the overall trends remain the same.

\subsubsection{Massive Clusters at Redshift 2}
\label{ooi}
Unlike the density defining population, which is selected based on stellar mass, we select our cluster of interest based on the halo mass - we simply take the most massive bright cluster galaxy (BCG) in the model at $z = 2$, which has a halo mass of $M_{halo} = 1.5 \times 10^{14}$ \Msun ~and a stellar mass of $M_{stars} = 1.86 \times 10^{11}$ \Msun. Our results were similar across several of the largest halos. There are 50 BCGs with $M_{halo} \ge 7 \times 10^{13}$ \Msun, which becomes $M_{halo} \ge 1\times 10^{14} M_{\odot}$ ~if $h = 0.7$. We define a BCG as the central galaxy of a cluster-sized halo \citep[see, e.g.,][]{deLucia2007}.

Our selection by mass is a luxury of using a simulation, as observers cannot do the same. Since our goal is not to mimic every observational detail, but rather to compare environment measures for a given cluster under specific, controlled selection circumstances, halo mass is one of the most obvious and simple selection criteria.

Note that an object of interest itself may or may not be a member of the density defining population. For the current work the former will always be true (since we focus on massive systems who always satisfy our density defining population criteria). However for observational data one may select the density defining population to be some volume limited sub-set of galaxies, for example, against which a magnitude limited population is analysed \citep[e.g.][]{Croton2005}.

\subsection{Environment Measures\label{env}}
With the density defining population in hand we now have a set of tracer ``particles'' (i.e. galaxies) with which to define environment at any point in the simulation box. But how does one define environment exactly? As there is no single agreed definition, we instead consider a range of metrics. As \cite{Muldrew2012} showed, none of the popular environment definitions currently employed in the literature probe the background density field in the same way, and the method used can colour an analysis in unexpected ways. It cannot be understated that these differences between environment methods must be understood if meaningful comparisons are to be made.

We select halo mass as our baseline measure of environment because the two are quite well correlated \citep[e.g.][]{Haas2012, Muldrew2012}, essentially fixing the mass and looking at the scatter in the environment for halos with $M_{halo} \ge 7 \times 10^{13}$ \Msun. In this paper we focus on the two most common categories of environment measure, the Nth nearest neighbour and the number of galaxies within a fixed aperture. Both are considered in three dimensions and in projection. Below we discuss each in turn and their application to our analysis.

\subsubsection{Nearest Neighbour Environments}
\label{NN}
The Nth Nearest Neighbour (NN) method is a simple concept but can become computationally intensive for higher values of N. In its basic form, the algorithm measures the distance to the Nth nearest density defining galaxy in two or three dimensional space around a chosen galaxy, with different authors adopting different values of N, typically 10 or fewer. The nearest neighbour distance can be translated into densities and then normalised by the mean density of the box. Here, we consider the central galaxy as N=0, although others sometimes count it as N=1.

Following \cite{Muldrew2012}, we define the 3D (volume) and 2D (surface) nearest neighbour densities around a galaxy by
\begin{equation}
\rho^{3D}_{NN} = \frac{N}{(4/3)\pi r^3_N}~
\end{equation}
and
\begin{equation}
\rho^{2D}_{NN} = \frac{N}{\pi r^2_N}~,
\end{equation}
respectively, where $r_N$ is the 2D or 3D distance to the galaxy being studied. The nearest neighbour density contrast is then given by
\begin{equation}
\delta_{NN} = \frac{\rho}{\bar{\rho}} - 1~.
\end{equation}
Here, the mean density of galaxies $\bar{\rho}$ is either the mean density of galaxies across the entire volume in the 3D case, labeled $\bar{\rho}^{3D}$, or for 2D is defined as
\begin{equation}
\bar{\rho}^{2D}=\bar{\rho}^{3D}\frac{2 v_{cut}}{H_0}~,
\end{equation}
where $v_{cut}$ is a chosen recession velocity cut (i.e. redshift) around the galaxy (with the factor of two measuring the volume in front of and behind the galaxy), and $H_0$ is the standard Hubble constant.

For all values of N the distribution of \dn\ is approximately lognormal and skewed slightly in the over-dense direction. Larger values of N display a tighter distribution in $\delta$ than smaller N values, demonstrating a higher variation in small-scale clustering than on larger scales where the distribution becomes more homogeneous.

Importantly, the nearest neighbour method depends on a variable length scale to quantify local density. As discussed in \cite{Muldrew2012}, nearest neighbour environment measures are useful for probing the internal properties of halos.

\subsubsection{Fixed Aperture Environments}
\label{FA}
The Fixed Aperture (FA) method draws a sphere (3D) or cylinder (2D) around the galaxy of interest and counts the number of density defining population galaxies inside. In this work we will consider apertures ranging in radius from $r=2$ to $20$ \Mpc, although a more typical scale in the literature would focus on $2$ \Mpc\ to $8$ \Mpc. For small apertures sampling a finite point distribution, such as in a galaxy survey, the fixed aperture method can be quite noisy. Larger apertures probe larger scales and hence have better signal-to-noise. For example, \cite{Croton2005} found that $r=8$ \Mpc ~optimally balanced signal-to-noise with survey depth while fairly sampling environments covering clusters to voids in the 2dF Galaxy Redshift Survey.

From a measurement of N density defining galaxies within an aperture of radius $r$ around the galaxy of interest, the fixed aperture overdensity is defined as 
\begin{equation}
\delta_{FA} = \frac{N}{\bar{N}} - 1~, 
\end{equation}
where, for the 3D spherical case, the mean number $\bar{N}$ is 
\begin{equation}
\bar{N} = \bar{\rho}^{3D}\, (4/3)\pi r_{FA}^3~, 
\end{equation}
while, for a 2D projected cylinder, $\bar{N}$ is 
\begin{equation}
\bar{N} = \bar{\rho}^{3D}\, \frac{2 v_{cut}}{H_0} \pi r_{FA}^2~, 
\end{equation}
and $\frac{2 v_{cut}}{H_0} $ is the depth of the cylinder, as before.

Similarly to the nearest neighbour distributions, the fixed aperture distributions are lognormal. Smaller scale probes of environment have more extended and noisy distributions, whereas larger scale probes are narrower, reflecting the increasing homogeneity of the large-scale Universe.

Fixing the scale at which environment is probed has both advantages and disadvantages. A fixed scale can simplify the interpretation of the results (e.g. there is significantly less need to decipher differing scale-dependent physics across environments probed with the same measurement). However, as discussed above, sparse distributions and small apertures can face non-trivial signal-to-noise problems. The benefits and drawbacks of both environment methods are discussed in Section \ref{discussion}.

\subsection{Redshift}
\label{redshift}

\begin{figure*}  
  \begin{center}
    \includegraphics[width = 6.5 in]{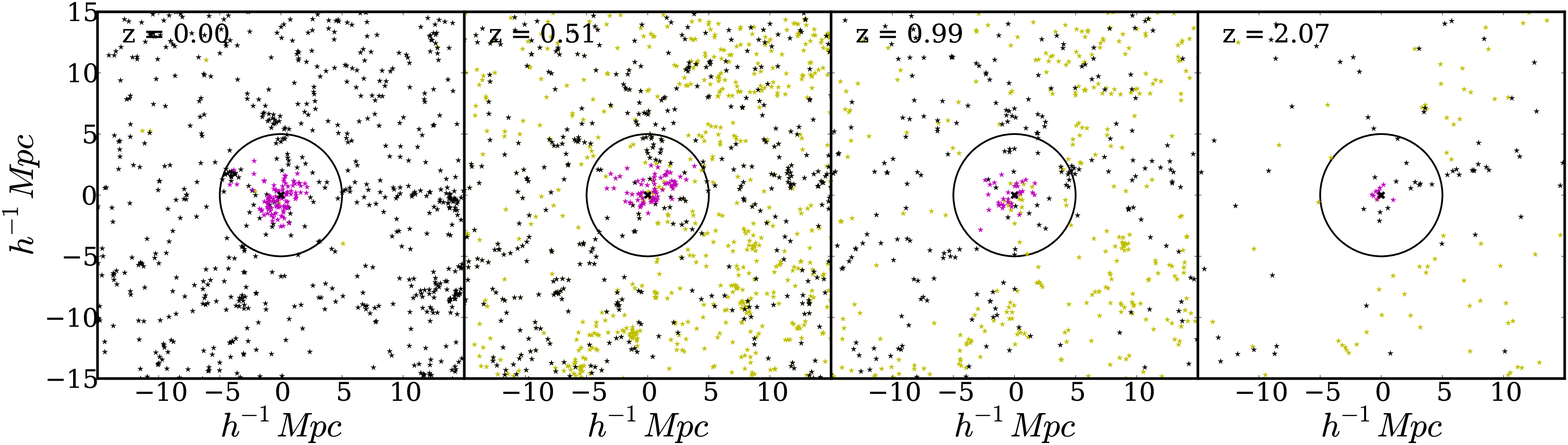}
  \end{center}
  \caption{The spatial distribution of galaxies around the largest cluster at $z = 2$ in our sample, projected onto a 30$\times$30 \Mpc\ area, as it appears at different epochs. The magenta points are bound to the background halo (the centre of which is marked by the black cross), the black points are galaxies within a spectroscopic-like velocity cut of $\pm 1000$\ \kms, and the yellow points are galaxies within a photometric-like velocity cuts of approximately $\Delta z/z \sim 0.1$. The circle indicates a 5 \Mpc ~projected aperture.\label{spatial}}
\end{figure*}

In observation, the accuracy of a galaxy's position is limited by its peculiar velocity along the line of sight. Our theoretical models allow us to ignore these perturbations, but we take them into account when measuring 2 dimensional, or projected, environment. These uncertainties dictate the depth of the cylindrical apertures and the velocity cutoffs of projected nearest neighbour calculations. Spectroscopic redshift cutoffs are determined by the velocity dispersion of the object under consideration. For a cluster this could be up to 1500 \kms. A peculiar velocity of this along the line of sight could easily move a galaxy into or out of a cluster, depending on the observer's point of view. Photometric redshift cutoffs are more common for high redshift studies, especially if statistically large samples are needed, because of the cost of getting enough spectra to make spectroscopic cuts. Photometric redshifts carry an uncertainty of $\frac{\Delta z}{z} \sim 0.1$, or $\Delta z\sim$\ 0.05, 0.1, and 0.2 at $z =$\ 0.5, 1, and 2, respectively. Since we are limited by the simulation box, we are slightly more generous in our photometric velocity cuts at high redshift, making them $\pm 15000$ \kms ($\pm 150$ \Mpc, using $\Delta d = v/H_0$) rather than $\pm 20000$ \kms\ ($\pm 200$ \Mpc). 

For simplicity, we ignore any additional error in redshift photometry might contribute and, in the local universe, we assume we have spectroscopic redshifts for all of the galaxies. This will mean that our photometrically sensitive results should contain more scatter than presented, and in this sense our analysis can be considered a conservative estimate of the truth. We have checked, however, that none of our conclusions change when this detail is implemented.

\section{Results}\label{results}

In this work we explore the environments of galaxy proto-clusters at high redshift. We quantify the accuracy to which such early-forming massive structures can be characterised in real observed samples. As environment is non-uniquely defined in the literature, we will focus on several popular methods and measure their uncertainties by ``observing'' massive galaxy clusters in our mock galaxy catalogue, constructed using the modified \citet{Croton2006} semi-analytic model, built upon the Millennium Simulation (see Section \ref{millennium}).

We begin with Figure \ref{spatial} ~which shows the changing environment around the most massive (at $z = 2$) simulated cluster in our mock catalogue in a projected 30 \Mpc\ area of co-moving volume, centred on the cluster,  at four epochs: $z=2$, $1$, $0.5$, and $0$. This cluster has a virial mass, defined as $200\rho_{crit}$, of $1.5$, $1.9$, $2.4$, and $6.5 \times 10^{14}$ \Msun\ at each of these redshifts, respectively. Small black (yellow) points mark the location of spectroscopic (photometric) density defining galaxies (Section \ref{ddp}), whereas magenta points mark actual cluster members identified in the semi-analytic model as bound to the halo. The outer circle indicates a radius of $5$ \Mpc\ around the cluster centre to calibrate the eye (this is a common scale over which environment is measured). This is considerably larger than the virial radius of the cluster dark matter halo, $R_{vir}$, which equals $0.47$, $0.65$, $0.83$, and $1.4$ \Mpc, respectively, and is too small to include in this figure.

\subsection{The most massive cluster at $z = 2$}\label{mostmassivez2}

\begin{figure*} 
  \centering
    \subfloat[]{\label{cylsph1}\includegraphics[width=3in]{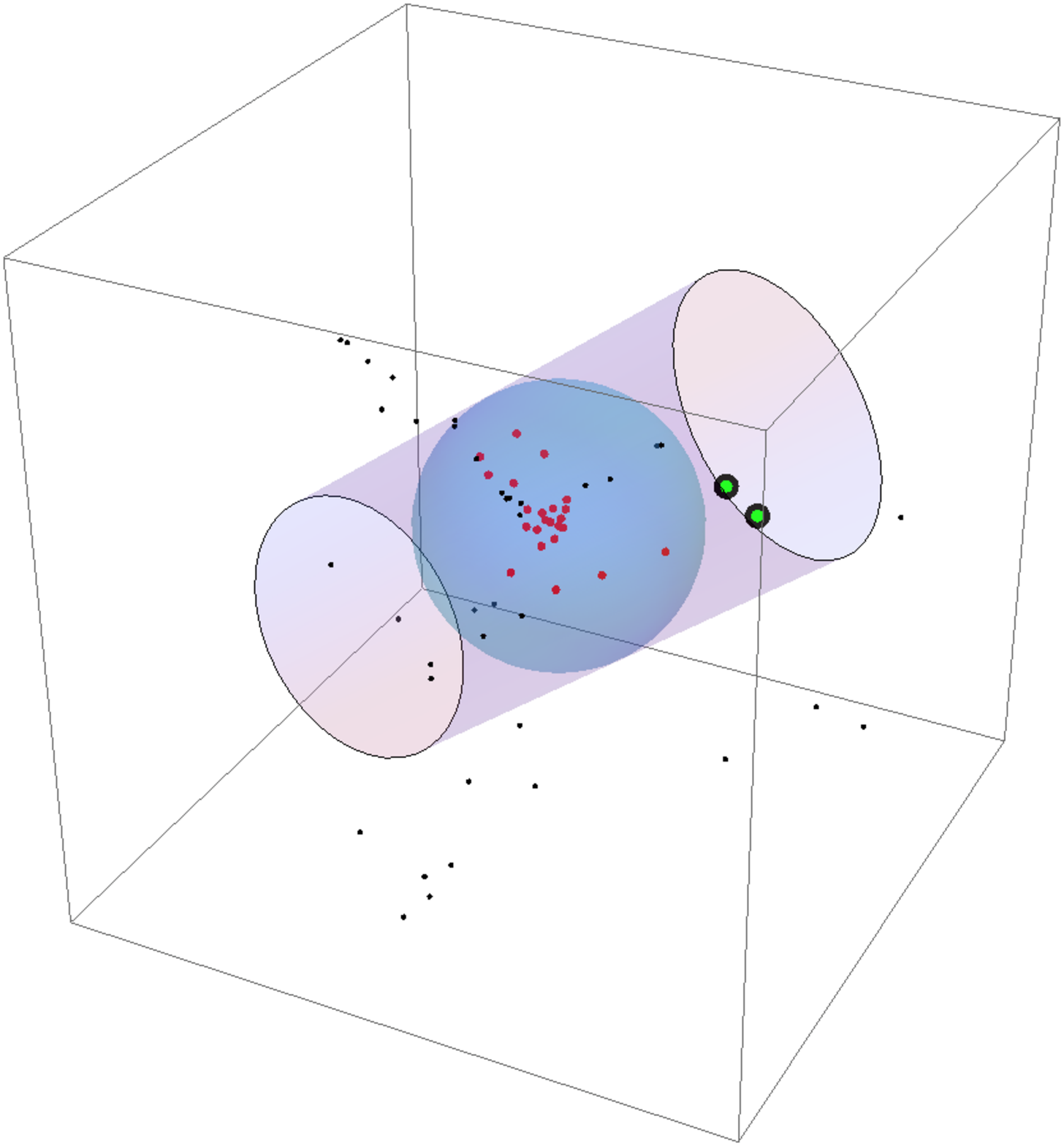}}
    \quad
    \subfloat[]{\label{cylsph2}\includegraphics[width=3in]{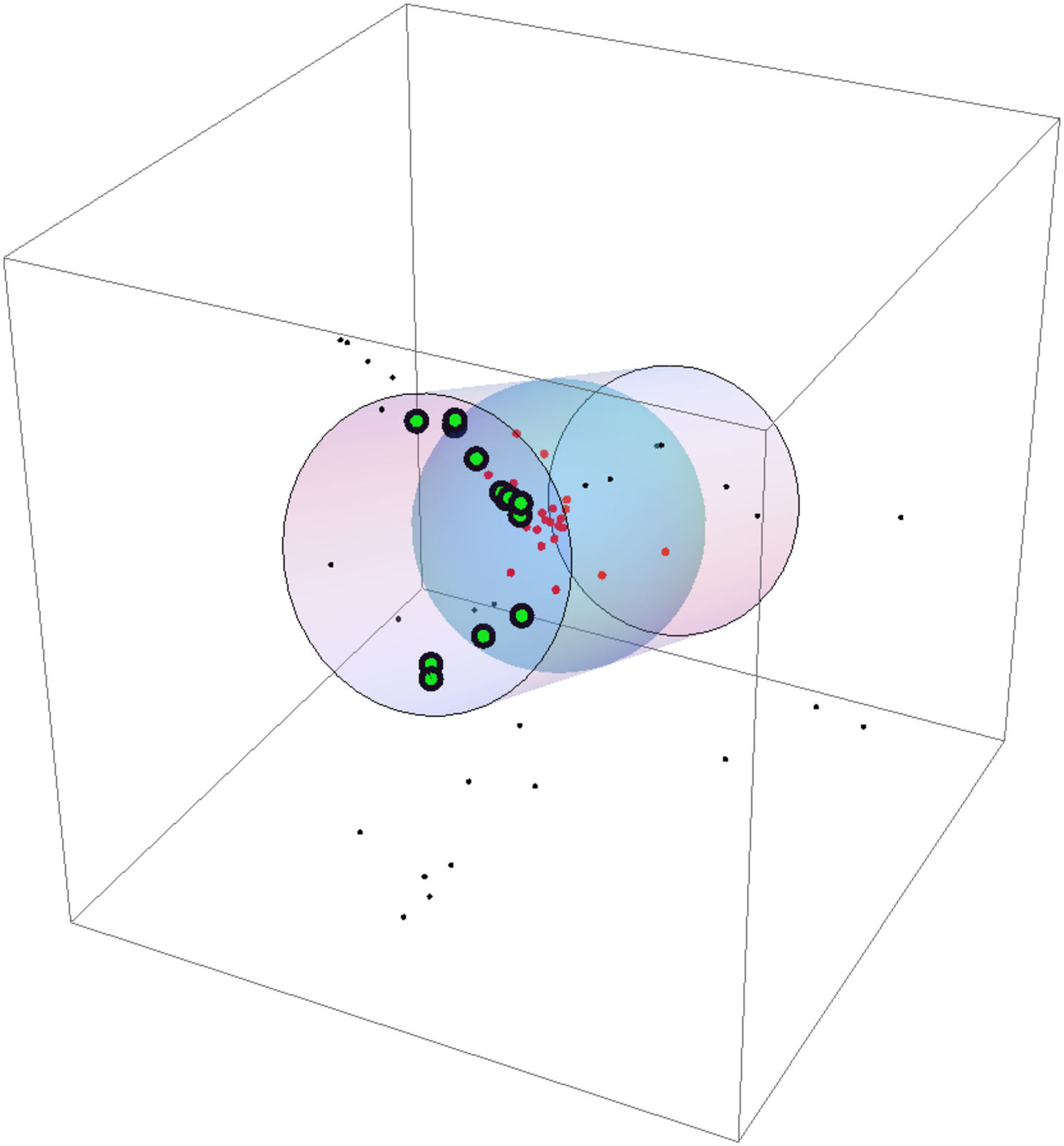}}
  \caption{The 3D real-space distributions of the galaxies in and near the most massive $z = 2$ cluster in our sample at two different viewing angles in a 25 \Mpc ~cube, centred on the largest galaxy. The red points are galaxies within the 3D spherical 5 \Mpc ~aperture. The green points ringed in black are the galaxies counted in the projected/cylindrical aperture but not the sphere. The black galaxies are outside both apertures. Figure \ref{cylsph1} has two green galaxies and Figure \ref{cylsph2} has thirteen.\label{cylsph}}
\end{figure*}

Figure \ref{cylsph}\ illustrates how such a cluster may be quantitatively probed by an overdensity metric. We show the three dimensional distribution of galaxies around the same cluster as in Figure \ref{spatial} at $z = 2$, centred on the halo centre. We superimpose two different kinds of fixed aperture probes on the galaxy distribution: a sphere and a spectroscopic cylinder of the same diameter. Red points mark galaxies which would be counted inside the sphere (and hence cylinder). Green points mark galaxies that are inside the cylinder but not the sphere. Black points show galaxies that are outside of both. The two panels are identical except the cylinder is oriented along a different viewing angle (and hence the galaxies within change). 

The green population in each panel highlights how a projected environment metric can be sensitive to the orientation of the cluster geometry and surrounding large-scale structure. Different viewing angles may result in significantly different number counts within the aperture, and hence significantly different quoted environmental over-densities, even though the galaxy distribution in all cases is identical. For example, in Figure \ref{cylsph} there is a $40\%$ change in the projected overdensity measurement between panels. For this example we have selected the most favourable conditions for the two environment measurements by taking them in real-space.

\begin{figure*}  
  \begin{center}
    \includegraphics[width = 6.5in]{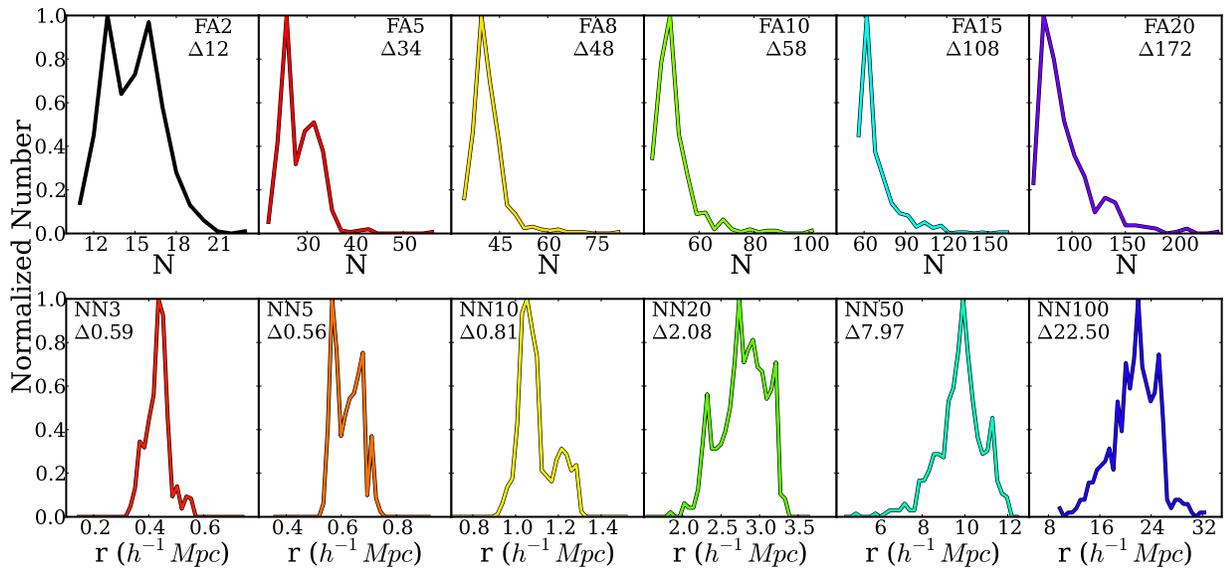}
  \end{center}
  \caption{Distribution of the number of galaxies measured around the largest cluster at $z = 2$ from 500 angles of observation using spectroscopic velocity cuts of 1000 \kms. We show a range of projected fixed aperture scales (top row) and distances to the Nth nearest neighbour for a range of N's (bottom row).  They are all arbitrarily normalized to the peak. The $\Delta$\ values are the difference between the highest and lowest measurement of N or r at each scale.  \label{var}}
\end{figure*}

\begin{figure}  
  \begin{center}
    \includegraphics[width = 3in]{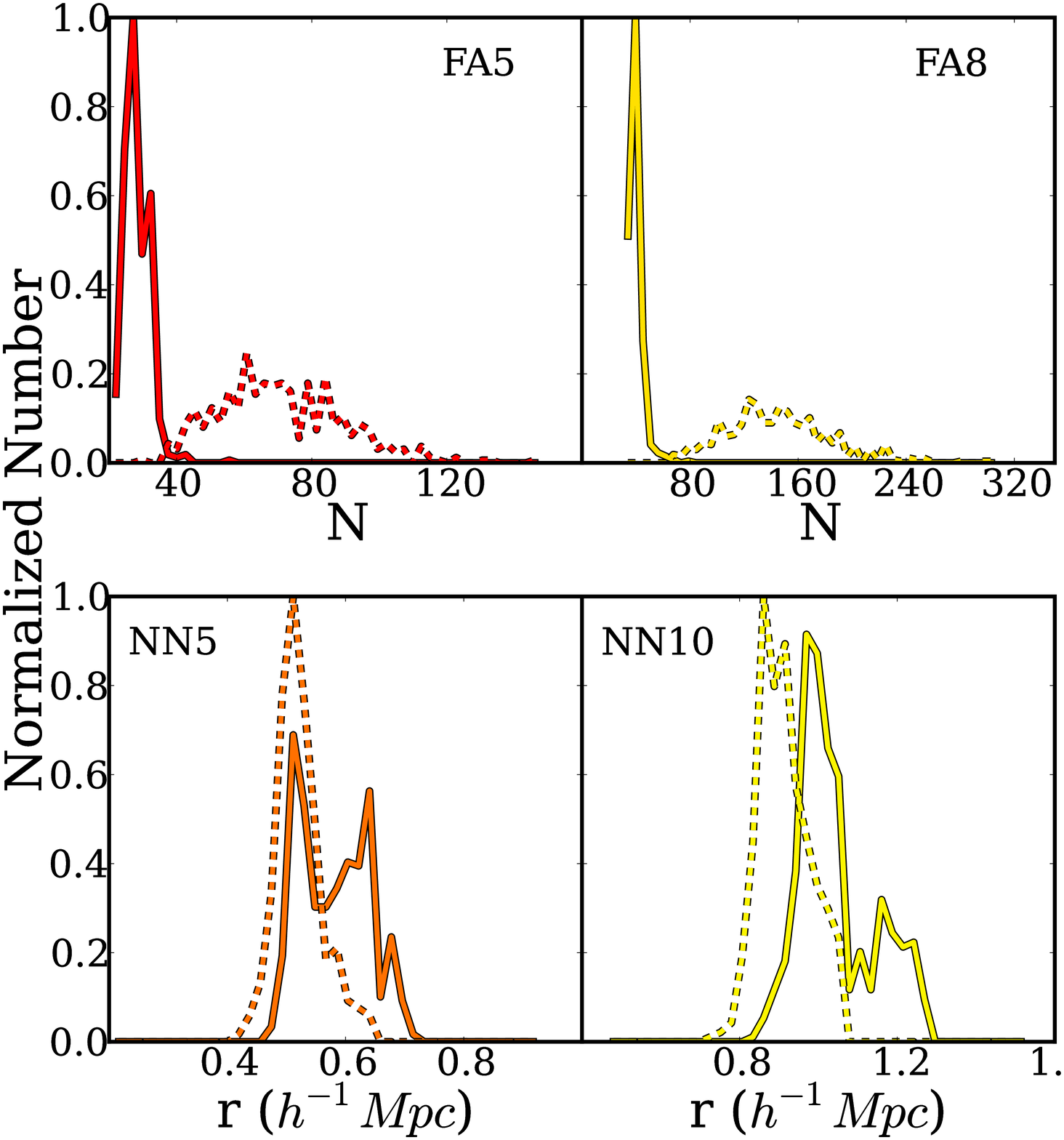}
  \end{center}
  \caption{Distribution of the number of galaxies measured in projected fixed aperture scales of $r_{FA}  = 5$\ and $8$ \Mpc\ (top row) and distances to the $5th$\ and $10th$\ nearest neighbour for a range of N's (bottom row) around the largest cluster at $z = 2$ from 500 angles of observation using both spectroscopic (solid lines - the same as the second and third columns in Figure \ref{var}) and photometric (dashed lines) velocity cuts of 1000 \kms and 15000 \kms, respectively. They are all arbitrarily normalised to the highest peak in the respective panel.\label{var_photo}}
\end{figure}

To explore this further we take the same system (still at $z=2$), view it from 500 random angles in redshift space and plot the distribution of projected galaxy counts, assuming spectroscopic velocity cuts centred on the halo centre. This is done for both environment metrics discussed in Section~\ref{method}: fixed aperture cylinders (as in Figure \ref{cylsph}) and nearest neighbour environment measures (which will also project differently on to the sky depending on viewing angle). The results are shown in Figure~\ref{var}. We consider a range of aperture sizes (top row) and nearest neighbour numbers (bottom row), as marked in each panel. The two peaks in the 5 \Mpc\ fixed aperture panel (FA5) come from finding filaments in the aperture. 

Figure~\ref{var} reveals the wide range of galaxy counts one may expect to find due to projection effects alone, and this wide range is independent of the environment metric used. Typically, larger scales are more singularly peaked, simply because the environment probe is large enough to smooth out the surrounding filaments very well. However, such filaments are the cause of the significant variations in galaxy counts seen at small scales of $r_{FA}$\ or small to intermediate values of N. Filaments are typically well defined features in the large-scale distribution, and hence a small change in viewing angle can cause a non-trivial number of galaxies to move in or out of the projected count. On the largest scales or for the largest N, one is sampling enough of the background density that the influence of individual filaments is lessened.

This smoothing is more apparent when photometric redshifts are considered. In Figure \ref{var_photo} we recompute the distribution of $r_{FA} = 5, 8$ \Mpc\ and N$ = 5, 10$\ with photometric cuts of $\pm \Delta z/z \sim 0.1$\ (dashed lines) and compare them to the same metrics from Figure \ref{var}\ (solid lines). Both fixed aperture examples show a vastly increased number of galaxies found inside the cylinder (which is to be expected with the much larger volume), as well as a more Gaussian distribution of counts. The 5th nearest neighbour photometric results are smoother than the 5th nearest neighbour spectroscopic results, but have a similar range as neither extends past 1.0 \Mpc. The radial distribution of the photometric 10th nearest neighbour measurements skews closer to 0.8 \Mpc\ than its spectroscopic counterpart, which skews closer to 1.0 \Mpc. Again, the photometric cuts lead to much smoother samples. These scales are where we see the effects of filaments along the line of sight in spectroscopic cuts but not photometric cuts. At larger scales, distributions using both metrics become far more Gaussian and shift to considerably higher values of N($< r_{FA}$) and lower values of $r_{NN}$(N).

\begin{figure}  
  \begin{center}
    \includegraphics[width = 3 in]{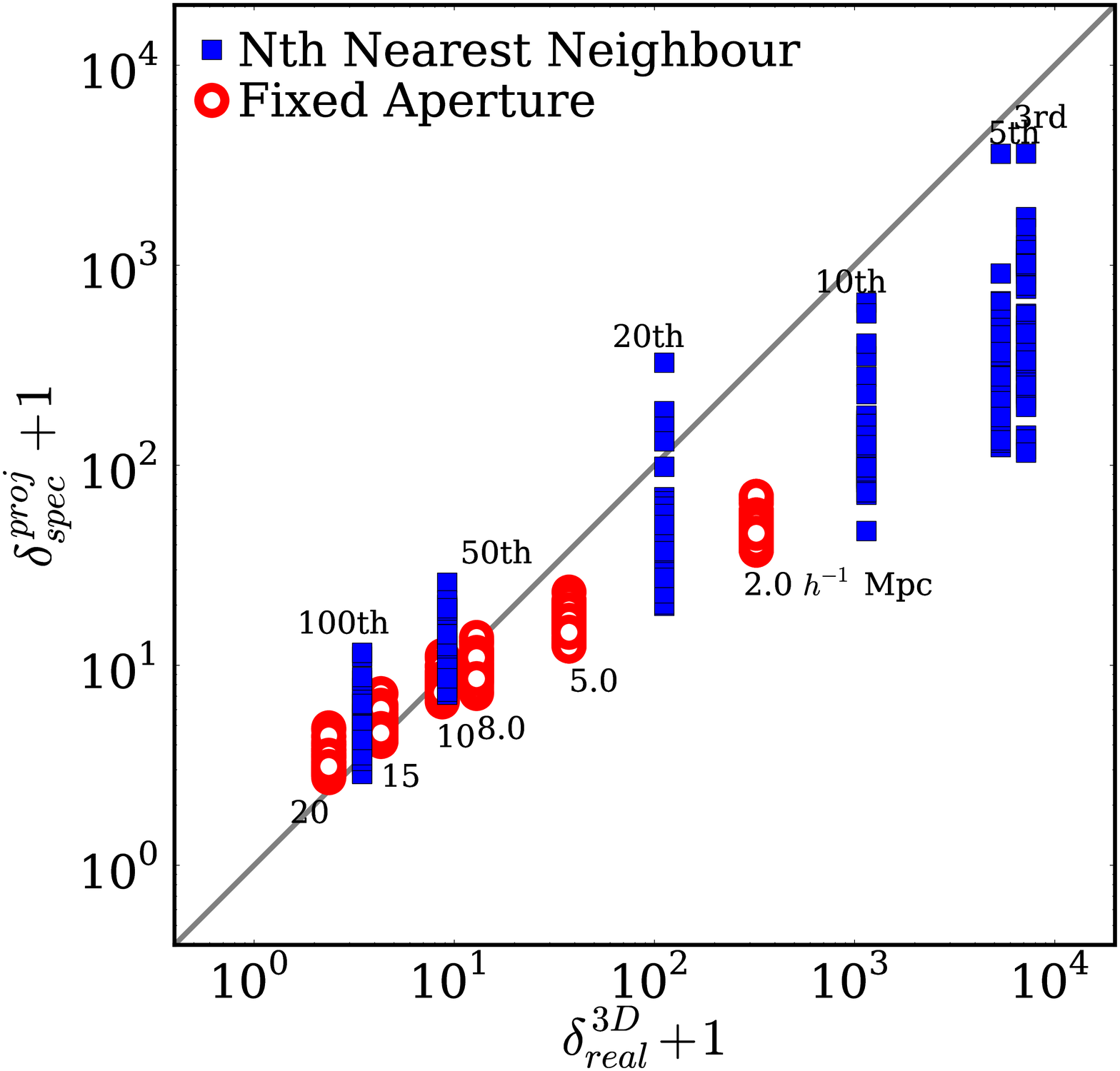}
  \end{center}
  \caption{The environment of the largest halo at $z = 2$, as viewed from 50 random angles. Here we compare ``observed" overdensities (\dps) in redshift space to ``actual" overdensities (\dtD) in real-space.\label{pov_z2}}
\end{figure}

It's often the case that a projected environment measurement is the best one can do given the limitations of the data on hand. In this case, one may like to know how well the projected density measurement recovers the actual three dimensional density that would be found with perfect data. This becomes especially important when interpreting the environment results. In Figure \ref{pov_z2} we compare each projected environment overdensity to its actual real-space three dimensional overdensity. This is again performed on the largest cluster at $z=2$. Since the projected overdensity is not unique but depends on the viewing angle, we measure 50 random angles with each environment metric. Nearest neighbour measurements are marked by blue squares, whereas fixed aperture measurements are marked with red circles. Each set of points span the collection of different viewing angles for the indicated aperture size or N.

We find an encouraging correspondence between 2D and 3D overdensity for all environment measures on large scales ($\ga 8$ \Mpc) and N ($\ga 10$), whereas smaller scale probes systematically under-predict the true environment in projection, and significantly so once the probe approaches the scale of individual halos. On the brighter side, and perhaps more importantly, the trends shown with changing scale or N are systematic, so relative behaviours in projection should be preserved in real three dimensional space.

Finally, similar to Figure \ref{var}, Figure~\ref{pov_z2} also quantifies the degree of scatter in the projected density measurements. Nearest neighbour projections appear much more volatile to projection effects than fixed aperture, where we see a significantly tighter relation. For N $\le$ 20, scatter in \dn\ is over 1 dex, compared to scatter in \df, which is less than 0.3 dex at all scales. For the same test using photometric cuts, we find scatter in the \df\ to be 0.5 dex or less at all scales and scatter in \dn\ to be very similar to our spectroscopic cuts shown here, although the 3D-2D correlation is not as well pronounced.

\subsection{The broader $z = 2$\ cluster population\label{Stats}}

So far we have focused on the most massive galaxy cluster at $z=2$ in our mock catalogue. We now expand this to consider a sample of 50 clusters at the same redshift, selected to have masses greater than $7\times 10^{13}$ \Msun. This will allow us to quantify the variance in our results and look for additional statistical features and environment metric trends.

\begin{figure*}  
  \begin{center}
    \includegraphics[width = 6.5 in]{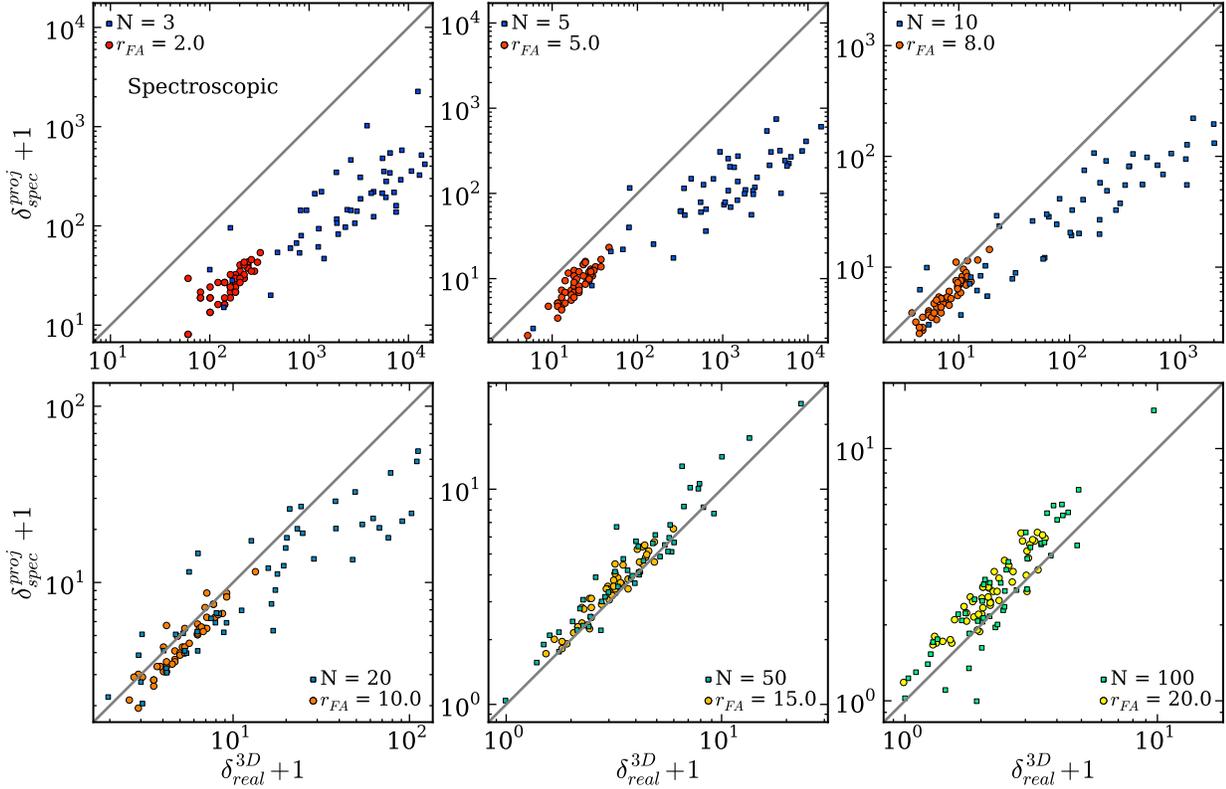}
  \end{center}
  \caption{The environment of the 50 largest halos ($M_{halo} \ge 7\times 10^{13}$ \Msun = $10^{14}$ ~M$_{\odot}$, assuming $h = 0.7$) at $z = 2$, comparing ``observed" overdensities (\dps) using spectroscopic cuts in redshift space to ``actual" overdensities (\dtD) in real-space. Squares are nearest neighbour measurements and circles are fixed aperture measurements.\label{gt1385_z2}}
\end{figure*}

Figure \ref{gt1385_z2} presents similar information as Figure \ref{pov_z2}, projected density vs. real-space three dimensional density, but now for the many environments of our cluster ensemble around the 50 largest haloes in the model. A different scale is represented in each panel, with the nearest neighbour and fixed aperture points having roughly equivalent scale lengths compared to the mean density. For each cluster, a random viewing angle was selected when measuring the projected environment.

Of note is the wide range of both projected and real environments that these massive clusters occupy, and this is shown by all environment measures. When comparing projected to real densities strong correlations surface. The systematic offset between the 3D and projected measurements stems partially from the different volumes (spherical and cylindrical) probed by the two metrics, and can therefore be easily adjusted, especially in the case of fixed aperture. This does not, however, account for the scatter in the projected measurements, which is similar in scale to the angle of observation scatter in Figure \ref{pov_z2}. At smaller scales, both fixed aperture and nearest neighbour methods consistently under-predict the true density when probed using the projected density. At all scales, nearest neighbour measurements show more scatter (and significantly so for small N).

\begin{figure*}  
  \begin{center}
    \includegraphics[width = 6.5 in]{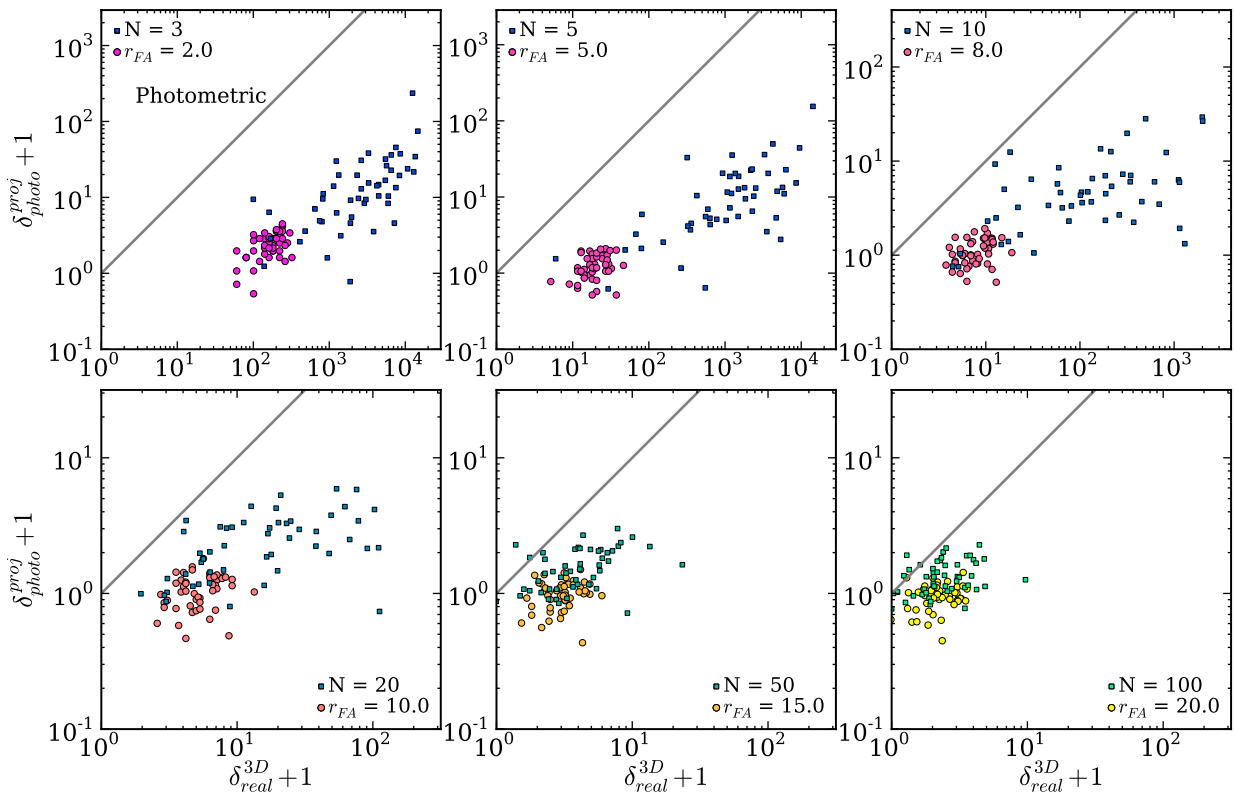}
  \end{center}
  \caption{The environment of the 50 largest halos ($M_{halo} \ge 7\times 10^{13}$ \Msun = $10^{14}$ ~M$_{\odot}$, assuming $h = 0.7$) at $z = 2$, comparing ``observed" overdensities (\dps) using photometric cuts in redshift space to ``actual" overdensities (\dtD) in real-space. Squares are nearest neighbour measurements and circles are fixed aperture measurements..\label{gt1385_z2_photo}}
\end{figure*}

Figure \ref{gt1385_z2_photo} shows the same comparison as Figure \ref{gt1385_z2}, but using photometric velocity cuts rather than spectroscopic. Similar to the spectroscopic example, at smaller scales the \dpp\ underestimates the \dtD, only by much more. Unlike the spectroscopic example of Figure \ref{gt1385_z2}, at larger scales, these galaxies' environments never converge on the 1:1 line, instead showing that a very large range of 3D real-space overdensities can be misidentified as a range of 2D photometric density contrasts. For example, a galaxy with a 3D 5th nearest neighbour overdensity of almost 1000 times the average density can be identified in projection as anywhere from 30 times the average surface density to one quarter of it. With a 5 \Mpc\ fixed aperture, almost $1/3$ of the galaxies are measured as inhabiting underdense regions, where only a few have 3D densities of less than 10 times the average. Similar tests with added uncertainty in the photometric redshift do not qualitatively change this figure.

\subsection{The Evolution of Environment} \label{evol}

One of the reasons high-redshift proto-clusters are so interesting is that they are expected to be the progenitors of the most massive and rare local galaxy clusters. Hence we are seeing them in their early stages of formation. Since proto-clusters are typically identified using the environment metrics discussed here, one may be curious to know exactly how this evolution unfolds with time. For example, how often are highly over-dense high redshift proto-clusters (as classified with some metric) still in such extreme environments by the present day? Or do they sometimes (often?) evolve in to something perhaps a little more mundane. In other words, how likely are you to find the progenitors of the most massive local clusters at high redshift using the environment metrics examined here?

To explore this we trace the galaxies and their parent dark matter halos in Figures \ref{gt1385_z2}\ and \ref{gt1385_z2_photo}~through time from $z=2$ to $z=0$. About 85\% of these galaxies survive as central galaxies at $z=0$ (the rest are subsumed into larger objects). We then calculate the same projected density measurements as performed at $z=2$ for each.

\begin{figure*} 
  \begin{center}
    \includegraphics[width = 6.5 in]{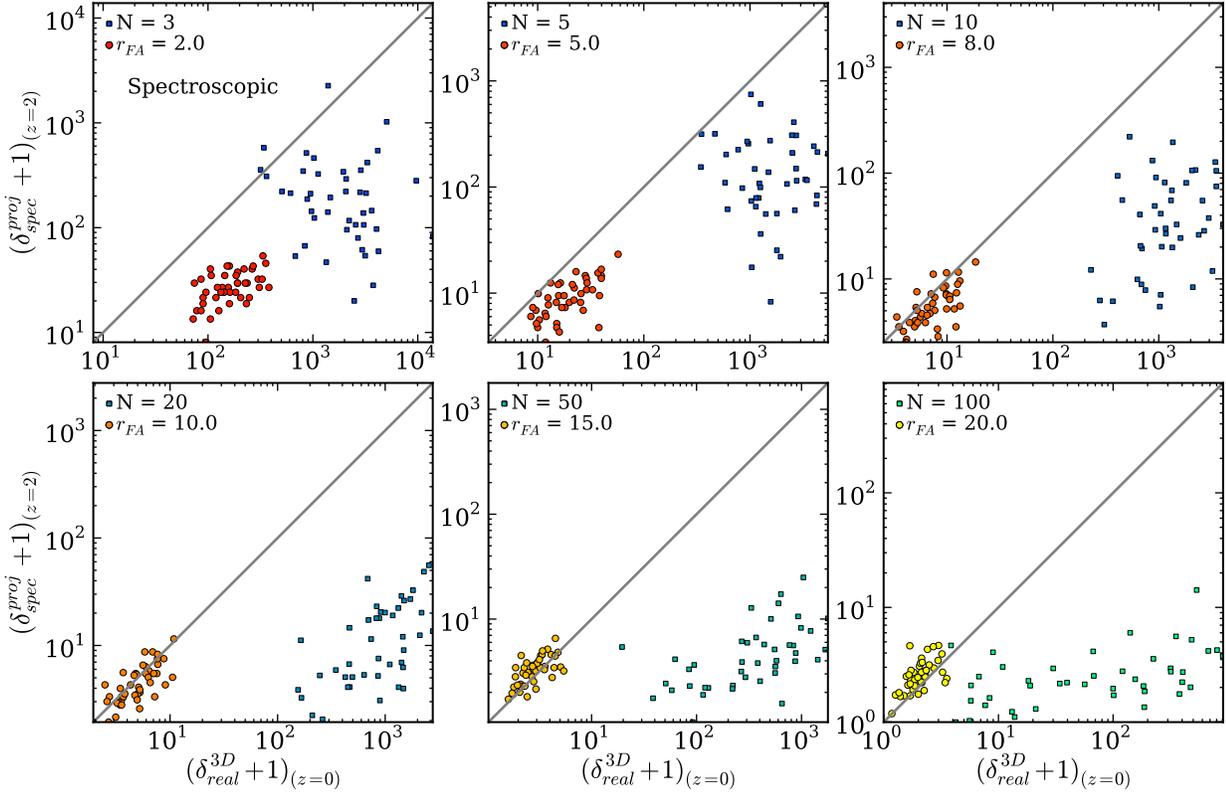}
  \end{center}
  \caption{The environment of the $\sim$40 largest halos ($M_{halo} \ge 7\times 10^{13}$ \Msun = $10^{14}$ ~M$_{\odot}$, assuming $h = 0.7$) at $z = 2$ that survive until $z = 0$, comparing ``observed" overdensities (\dps) using spectroscopic cuts in redshift space at $z = 2$ to ``actual" overdensities (\dtD) in real-space at $z = 0$. Squares are nearest neighbour measurements and circles are fixed aperture measurements.\label{z2-0}}
\end{figure*}

In Figure \ref{z2-0} we compare the projected over-densities for this sample of massive halos at $z=2$\ as would be observed in a spectroscopic sample, to the actual real-space overdensity each evolves into by $z=0$, to see how the ``observed" measurements at $z = 2$\ hold up as the cluster evolves. Squares indicate nearest neighbour measurements and circles the fixed aperture measurements. As in Figures \ref{gt1385_z2}\ and \ref{gt1385_z2_photo}, each panel contains both a nearest neighbour and fixed aperture metric which probe approximately the same scale, compared to $<r_{NN}>$\ and $<N_{FA}>$ calculated from the average density.

 At all scales, fixed aperture over-densities are reasonably well correlated between high and low redshift. Nearest neighbour over-densities, on the other hand,  show significant scatter, with the relatively tight correlation between 2D ($z = 2$) and 3D ($z = 2$) measurements from Figure \ref{gt1385_z2} disappearing almost entirely as the galaxy evolves to $z = 0$. Evolution can scatter the density  contrast by up to two orders of magnitude between $z = 2$ and $z = 0$. Comparisons between \dps$(z = 2)$ and \dps$(z = 0)$ as well as \dpp$(z = 2)$ and \dpp$(z = 0)$ show similar (lack of) patterns.


\begin{figure} 
  \begin{center}
    \includegraphics[width = 3 in]{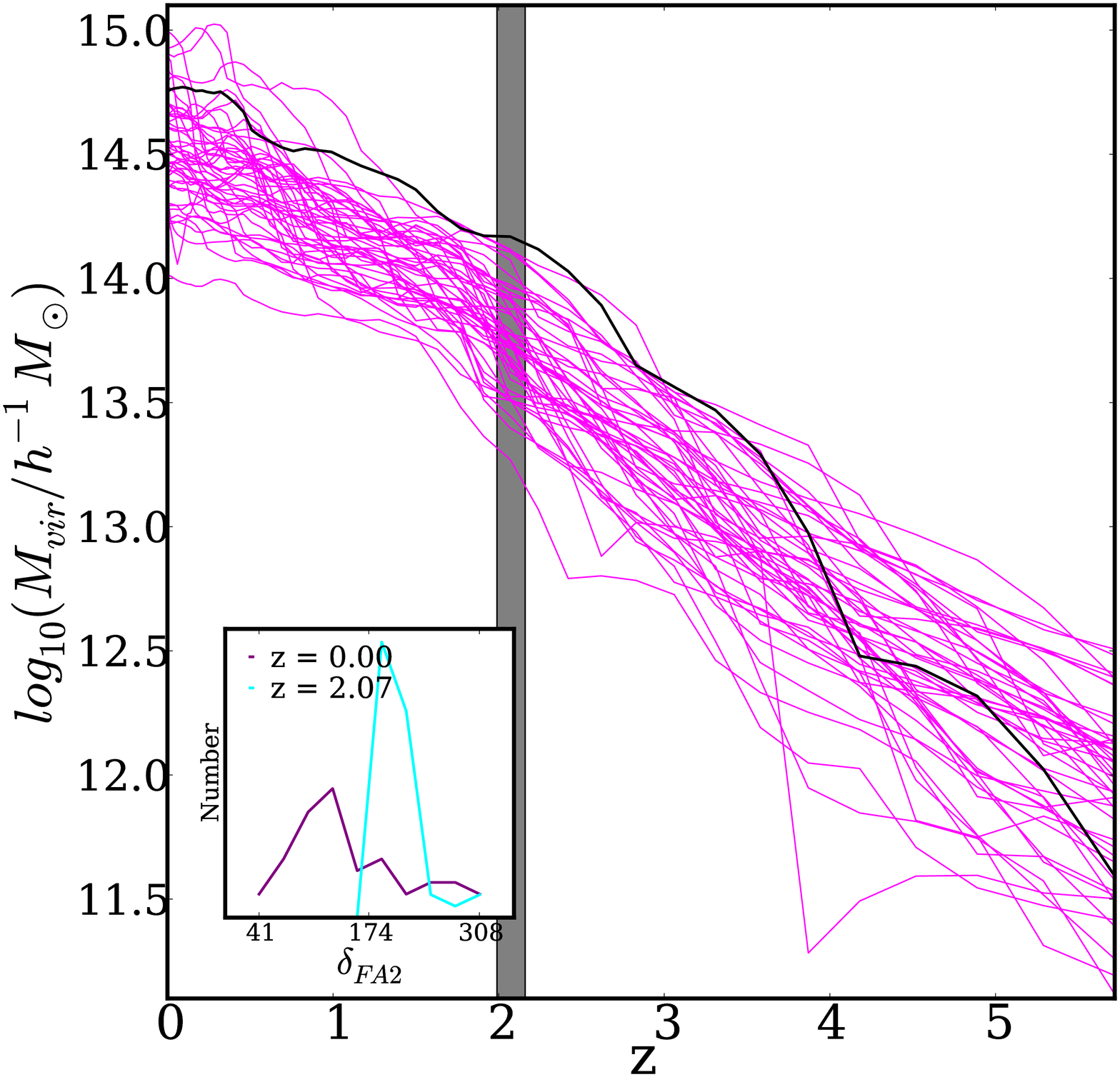}
  \end{center}
  \caption{The mass history of the 50 central galaxies with the highest small scale fixed aperture environment measure on a spherical 2 \Mpc\ scale, $\delta_{FA2}$, at $z = 2$, marked by the grey vertical band. Their parent halos are then are traced from $z = 6$\ to the present in the simulation, with the black line representing the history of the largest halo at $z = 2$\ (the halo considered in previous figures). The spread in halo masses is quite sizeable throughout the history, including at the point of selection at $z = 2$. \label{mass_fa2}}
\end{figure}

Thus far we have selected galaxies based on their halo masses, a luxury of using a simulation. For observed data, such properties are much harder to quantify, especially for large galaxy samples. Because of this, observers instead will select their objects of interest using redshift and environment \citep[e.g.][]{Spitler2012}, or redshift, environment, and colour \citep[e.g.][]{Papovich2008}. To mimic more closely this type of selection, in Figure \ref{mass_fa2} we select the 50 most dense galaxies at $z = 2$, as defined by $\delta_{FA2}$, the real-space spherical fixed aperture on a $2$ \Mpc\ scale. We ask what are the halo histories of such observationally over-dense high redshift galaxies? Are these objects always in the most massive halos? Are they still overdense at $z = 0$?

Figure \ref{mass_fa2} shows that there is a significant variation in the halo mass and evolution of such objects, of the order of 0.5-1.0 dex, and this is true at all redshifts probed. As a partial answer to the second question above, galaxies selected in this manner do not occupy the largest haloes at $z = 2$, leaving little likelihood that they will do so at $z = 0$. The inset panel highlights the change in the distribution of overdensities from our selected galaxies at $z = 2$ (cyan) to a larger spread of overdensities at $z = 0$ (purple). It is important to note that the increase in the spread of overdensities reflects the scatter of environments due to evolution of the clusters as well as the evolution of the density defining population, as discussed in section 2.2.1. Galaxies that have a specific overdensity at $z = 0$ have almost 10 times the number of galaxies in the aperture as galaxies with the same $\delta$\ at $z = 2$, e.g. while the density contrast is decreasing with decreasing redshift, the local density itself is increasing. Including colour is beyond the scope of this paper, but it is clear that something more than just environment is needed to accurately select the most massive proto-clusters at $z = 2$.

\section{Discussion and Conclusions}\label{discussion}

In Section \ref{Introduction} we asked three questions about the identification of high redshift proto-clusters using their environment as a probe. These were (1) how the 2D environment compares to the actual 3D environment using different metrics, (2) for a single metric, how viewing angle changes the measured 2D environment, and (3) how stable a particular metric is with time.
 
Figures \ref{cylsph} -- \ref{z2-0} show that the difference between ``observed" overdensity and ``actual" overdensity is sensitive to both scale and method. For example, on smaller scales the potential for large deviations from the actual environment increases for both nearest neighbour and fixed aperture methods, as seen in Figure \ref{gt1385_z2}. On larger scales (i.e. larger than r=20 \Mpc\ or N=50) the agreement improves, but this is primarily because the Universe is more homogeneous. Not surprisingly, we see a break down in the correlation when using photometric velocity cuts.

When comparing the two environment metrics at similar scales we find different sensitivities: nearest neighbour tends to have more scatter in the 2D vs 3D overdensity whereas fixed aperture tends to have less. Also of note, at all but the largest scales, the scatter in the fixed aperture measurements does not increase as much as the scatter in the nearest neighbour measurements when switching to photometric redshift cuts from spectroscopic redshift cuts. Figures \ref{gt1385_z2}\ and \ref{gt1385_z2_photo}\ are broken down by scale, so it's easy to see the variation from scale to scale as well as method to method.

As expected, the scatter in $\delta^{proj}$ increases dramatically between the spectroscopic-like (Figure \ref{gt1385_z2}) and photometric-like (Figure \ref{gt1385_z2_photo}) cuts. There is almost no correlation between \dpp\ and \dtD\ measurements for N $\ga 10$ or $r_{FA} \ga 8$ \Mpc. This is not quite as dire as \cite{Cooper2005} suggest, but does imply a caveat for environment studies done using large apertures or values of N. The aperture size or value of N where the correlation disappears is highly dependent on the density defining population, so spectroscopic redshifts are vital for accurate environment measurements (angle of observation variations aside).

A key hindrance to measuring environment in projection is that the angle of observation matters quite a bit -- a galaxy measured straight down a filament has a very different projected overdensity than that of a galaxy measured perpendicular to the filament. Figures \ref{cylsph} - \ref{pov_z2} highlight the minimum certainty with which a projected overdensity can be ``known". Interestingly, the angle of observation has a much larger impact on nearest neighbour measurements than it does on fixed aperture measurements, as seen in Figure \ref{pov_z2}, although nearest neighbour measurements are less affected by the uncertainty resulting from photometric measurements.

In Figure \ref{z2-0} we test the stability of the metrics across time. This is not only to measure the evolution of the environment but also to link these rare $z = 2$ objects to their better-studied counterparts in the local universe, i.e. we are now comparing the observed environments at high redshift to actual environments of current objects. Figure \ref{z2-0} shows that for all N, the actual nearest neighbour overdensities of the cluster galaxies are larger at $z = 0$ than the observed values at $z = 2$, with the exception of three galaxies measured with $N = 3$. This is expected from hierarchical structure formation where objects come together with time via gravitational instability. For example,  for $N \le 10$, the environments of clusters covering a 2-3 dex range of overdensity at $z = 2$ all evolve to become very overdense at $z = 0$ with about half the spread in dex. However, measurements  using $N > 20$ show the opposite relation - overdensities no more than 1 dex from the mean at $z = 2$ evolve into environments that span 2-3 orders of magnitude at $z = 0$. Since we have an order of magnitude more galaxies in our density defining population at $z = 0$\ than at $z = 2$, the value of N used to probe overdensities at equivalent scales changes between the two redshifts.

Fixed aperture measurements are much more consistent in their evolution from $z=2$\ to $z=0$, are much closer to the 1:1 line, and have a much smaller spread. There is slightly more scatter on a 2 \Mpc\ scale, and the population shifts from just below the 1:1 line to just above it as the scale increases. Since the background density is increasing, the density around the galaxy is also increasing, just not at quite the same pace at all scales. We find that a single galaxy's fixed aperture environment measures are very stable across time. The relative consistency of the fixed aperture measurements might be a matter of the inner galaxies of clusters falling in at a rate faster than galaxies on the outside, so on a larger-than-cluster scale (which includes all of the fixed aperture scales in this paper), the relative density does not change much at that scale, but the core of the cluster (and therefore the nearest neighbour measurements) becomes much denser relative to the past.

From this work, we can say that of the several large recently discovered clusters and proto-clusters (as discussed in Section \ref{Introduction}), many, if not all are likely to remain or become clusters at $z = 0$, although they might not still be the largest, as shown in Figure \ref{mass_fa2}. In fact, many may end up as more run-of-the-mill low redshift cluster objects. That said, in almost all of the simulated examples in Figure 7, the actual 3D overdensity is larger than the projected overdensity, indicating that the recently discovered clusters probably have a higher density contrast than observed.

Additionally, \citet{Diener2013} make use of mock catalogues based on the zCosmos survey \citep{Lilly2007} to find the likelihood high redshift ($1.8 < z< 3$) groups and proto-groups will evolve into large low redshift clusters. Selecting by environment rather than halo mass, they find they should have detected $65\%$ of the progenitors to today's massive clusters, assuming a complete survey. This is a similar result to our Figure \ref{mass_fa2}.

To support the identification of a high redshift proto-cluster candidate, observers will often apply an additional colour restriction to their cluster sample to single out red-sequence galaxies that are characteristic of such massive objects at all redshifts \citep{Papovich2010}. While this should certainly add confidence in the reality of a particular proto-cluster candidate, we note that high redshift, highly over-dense massive galaxies are also characterised by their significant star formation and blue colours \citep{Cooper2007, Elbaz2007}, unlike at low redshift. Hence, applying such a colour constraint may mean that many proto-clusters are missed in the observations. For the present work we do not use colour and magnitude in our selection; such observables are subject to significant uncertainties and their accurate modelling is difficult at these redshifts. This will be addressed in future work.

\section*{Acknowledgments}
GS is supported by a Swinburne University SUPRA postgraduate scholarship. DC acknowledges the receipt of a QEII Fellowship from the Australian Research Council. RAS is supported by the NSF grant AST-1055081. We thank the referee for their constructive comments and attention to detail.

The Millennium Simulation used in this paper was carried out by the Virgo Supercomputing Consortium at the Computing Centre of the Max-Planck Society. Semi-analytic galaxy catalogs from the simulation are publicly available at http://www.mpa-garching.mpg.de/millennium/. 

The mock galaxy catalogue generated from the Millennium Simulation trees can be downloaded from the Theoretical Astrophysical Observatory (TAO) repository: http://tao.it.swin.edu.au/mock- galaxy-factory/.

\bibliographystyle{mn2e}
\bibliography{refs}


\end{document}